\documentclass[11pt]{article}

\usepackage{authblk}

\begin{document}

\title{Unwrapping Chains}

\author[1]{A. D. Cambou}
\author[1]{B. D. Gamari}
\author[2]{E. Hamm}
\author[1]{J. A. Hanna\thanks{hanna@physics.umass.edu}}
\author[1]{N. Menon}
\author[1]{C. D. Santangelo}
\author[1]{L. Walsh}
\affil[1]{Department of Physics, University of Massachusetts, Hasbrouck Lab, 666 N. Pleasant St., Amherst, MA 01003, USA.}
\affil[2]{Departamento de F\'{i}sica, Universidad de Santiago de Chile, av.\ Ecuador 3493, 9170124 Estaci\'{o}n Central, Santiago de Chile, Chile.}

\date{\today}

\maketitle

\begin{abstract}
A loop of chain can move along its own tangents, maintaining a steady shape.  An open-ended chain undergoing a nontrivial motion must change its shape.  One consequence is that chains pulled around objects will fail to follow the contours of the objects, unwrapping themselves instead.  This short note accompanies a fluid dynamics video submission (83068) to the APS DFD Gallery of Fluid Motion 2012.
\end{abstract}

The accompanying video shows chain (stainless steel, 3.3 mm flat oval cable, {\footnotesize{\texttt{Rio Grande, Albuquerque, NM}}}) unwrapping from various obstacles, as one end is pulled in an approximately straight line at speeds on the order of 2-3 m/s.

While other forces, including gravity and friction, are operative in these demonstrations, the behavior of the free end may be explained by the interplay of inertia and a line tension that serves to locally preserve arc length.  These two elements combine in the string equations \cite{Routh55}, which admit steady loop configurations with uniform positive tension, the material points moving purely along the tangents of a curve of steady but arbitrary shape.  The tension is analogous to a volume-preserving pressure in a one-dimensional fluid.  It must fall to zero at a free end.  Gradients in this tension generically imply a time dependence of the curvature of the chain.  The result is that a chain with a free end cannot maintain a steady shape.  Though its initial trajectory may follow the contours of an obstacle, it will subsequently try to squeeze the obstacle or unwrap.  

Also apparent in some of these films is the whip-like motion of the chain, when the waveform corresponding to the obstacle reflects from the free end.

%
%
%

\section*{Acknowledgments}

We thank J.-C. G\'{e}minard for introducing us to this phenomenon, and for very enlightening discussions.  We also thank P.-T. Brun and D. Vella for sharing results of their work with us.  We acknowledge funding from U.\ S.\ National Science Foundation grants DMR 1207778, DMR 0907245, and DMR 0846582.

\bibliographystyle{unsrt}

\end{document}